\documentstyle[aps,manuscript]{revtex}

\begin{document}

\title {Crowd-anticrowd model of the Minority Game}
\author{M. Hart, P. Jefferies and N.F. Johnson}
\address {Physics Department, Oxford University,
Oxford, OX1 3PU, U.K.}
\author{P.M. Hui}
\address {Department of Physics, The Chinese
University of Hong Kong, Shatin, \\
New Territories, Hong Kong}

\maketitle

\begin{abstract}
We provide a theoretical description of the  
Minority Game in terms of crowd effects. The size of the
fluctuations arising in the game is controlled by the interplay between crowds of
like-minded agents and their anti-correlated partners (anticrowds). The
theoretical results are in  good agreement with the numerical
simulations
over the entire parameter range of interest.

\end{abstract}
\bigskip


\newpage

The study of agent-based models of complex adaptive systems
is attracting much attention\cite{arthur}. Among the many possible
interdisciplinary applications is the growing field of
econophysics\cite{stanley}: each agent knows the past ups and downs in the
index
of a financial market and must decide how to trade based on this global
information. The Minority Game (MG) introduced by
Challet and Zhang\cite{challet,savit}, offers arguably the simplest paradigm
for
such a complex, adaptive system.  The MG comprises an odd number
$N$ of agents, each equipped with $s$ strategies and a memory size
$m$, who repeatedly compete to be in the
minority\cite{challet,savit,spinglass,sherrington,dHR,cavagna,us,crowd}.
The most
striking feature arising from numerical simulations is the
non-monotonic variation in the size of the fluctuations (i.e. standard
deviation) produced by the MG as
$m$ is varied\cite{savit}.
Challet {\em et al}
provided a sophisticated formal connection between the MG and spin
glass systems\cite{spinglass} which offers many fascinating quantitative
insights into the MG's dynamics. Given the 
complexity of this dynamics, it is
understandable that no general theory has yet been proposed which yields
quantitative agreement with the numerical results of Ref. \cite{savit} over
the
full range of
$m$, $s$ and $N$ values.

In this paper, we show that a theoretical
model can be constructed in a surprisingly simple way by incorporating
the `crowd' effects (i.e. strong inter-agent correlations) which arise
within the
interacting, many-agent population. The results yield good agreement
with the numerical results\cite{savit} 
over the entire range of $m$, $s$ and $N$.
The
non-monotonic behaviour of the standard deviation\cite{challet,savit} 
is shown to
arise from a fascinating interplay between a crowd and its anti-correlated
partner
(`anticrowd').

The MG\cite{challet} comprises an
odd number of agents $N$ who choose repeatedly between
option 0 (e.g. buy) and option 1 (e.g. sell).   
The winners are those in the
minority group, e.g. sellers win if there is an excess of buyers. The
outcome at
each timestep represents the winning decision, 0 or 1. 
A common bit-string of the $m$ most recent
outcomes is made available to the agents at each timestep.
The agents
randomly pick $s$ strategies at the beginning of the game, with repetitions
allowed. After each turn, the agent assigns one (virtual) point to each of his
strategies which would have
predicted the correct outcome. At each turn of the game, the agent
uses the most successful strategy, i.e. the one with the 
most virtual points, 
among his $s$ strategies. The strategy-space ${\cal V}_m$ forms a
$2^m$-dimensional  hypercube with
strategies at the $2^{2^m}$ vertices. Fortunately,
the MG's standard deviation is essentially unchanged if a `reduced' strategy
space ${\cal U}_m$ is used instead of ${\cal V}_m$ \cite{challet}: the
${\cal U}_m$ only contains
$2^{m+1}$ strategies or equivalently $2^m$ strategy pairs $\{\cal G\}$.
The two strategies within a given pair $\cal G$ are anticorrelated, i.e. they
differ by the maximum possible Hamming distance $d_H=2^m$ \cite{challet}.
Strategies between any two pairs
$\cal G$ and $\cal G'$ are uncorrelated, i.e. they differ by $d_H=2^{m-1}$.  The
results presented in this paper employ the reduced strategy space. 

If ${n}_R$ agents use the same strategy $R$, then they
will act as a `crowd', i.e. they will make the same decision. If
${n}_{\bar R}$ agents simultaneously use the 
anticorrelated strategy  
$\bar R$, they will make the opposite decision and will hence act as an
`anticrowd' (${\cal G}\equiv (R,\bar R)$). If
${n}_R\approx {n}_{\bar R}$ for all $\cal G$, then the actions of the crowds and
anticrowds cancel and the standard deviation $\sigma$ of the number of agents
choosing a given option (the so-called `attendance' time-series
$A(t)$) will be small. In contrast if
${n}_R
\gg n_{\bar R}$ for all
$\cal G$, then
$\sigma$ will be large. Since there is no correlation between $\cal G$ and
$\cal
G'$, each group $\cal G$ comprising a
crowd-anticrowd pair $(n_R,n_{\bar R})$ will contribute to the attendance
$A(t)$ via a separate random walk in time of step-size
$|n_R-n_{\bar R}|$. The variances of these walks can then be summed to
obtain the standard deviation of $A(t)$. Hence we describe the MG standard
deviation using the following theoretical expression:
\begin{equation}
\sigma = \bigg[\sum_{{\cal G}=1}^{2^{m}} \sigma^2_{\cal
G}\bigg]^{\frac{1}{2}} =
\bigg[\sum_{{\cal G}\equiv (R,{\bar R})}
\frac{1}{4}|n_R-n_{\bar R}|^2\bigg]^{\frac{1}{2}}
\
\end{equation}
where both time-averaging, for a given configuration of
initial strategies, and configuration-averaging have been
carried out. We now demonstrate that this crowd-anticrowd cancellation underlies
the numerical results for
$\sigma$ vs. $m$
\cite{savit}.
We run the numerical simulation of the MG and wait until transients in
$A(t)$
have disappeared. At timestep $t_0$, we read out the number of players
playing each strategy $R$, where $R=1,2,\dots 2^{m+1}$. For each strategy pair
${\cal G}=(R,{\bar R})$, we calculate 
$n_R-n_{\bar R}$ at time $t_0$ and hence obtain $\sigma$.  We then
average this $\sigma$ over 1000 timesteps to simulate the time-averaging.
We have checked that our
results are insensitive to the precise time-averaging procedure.  Finally,
we
average over 16 runs to simulate the configuration-averaging.
Figure 1 compares the resulting time and configuration-averaged $\sigma$ with that
obtained from the numerical MG simulation. The agreement is very good for all
$m$, $s$ and
$N$ (not shown). We conclude that the crowd-anticrowd cancellation can
indeed quantitatively explain the numerical results of Ref. \cite{savit}.

We need expressions for the number of agents using each strategy, i.e.
$\{n_R\}$.
Since the labels $R$ are arbitrary in Eq. (1), the ordering of
strategies
$\{n_R\}$ has no particular significance.
At any particular time
$t_0$, these
$2^{m+1}$ strategies can be ranked according to their virtual points by a
sort-operation $\Theta$ acting on the list $\{n_R\}$. Hence
$\{n_R\}\stackrel{\Theta}\mapsto \{{n}_\rho\}$ where $\rho$ is
the virtual-point rank label with $\rho=1$ being the highest scoring
strategy and
$\{n_\rho\}\equiv n_{\rho=1}, n_{\rho=2},\dots n_{\rho={2^{m+1}}}$. Another
useful
method of ordering is achieved by ranking strategies $\{n_R\}$ at time
$t_0$ according to their
popularity. In particular, strategy $r=1$ is defined as the strategy
which is being used 
by the largest number of agents, strategy $r=2$ is being used
by the second largest number of agents, etc. We denote the 
popularity ranking by
$\{n_r\}$ where
$\{n_r\}\equiv n_{r=1}, n_{r=2},\dots n_{r={2^{m+1}}}$. 
Note that $\{n_r\}$ can be
obtained from $\{n_R\}$ and hence $\{n_\rho\}$ by sort operations, i.e.
$\{n_R\}\stackrel{\Psi}\mapsto \{n_r\}$ and
hence $\{n_\rho\}\stackrel{\Gamma}\mapsto
\{n_r\}$.
Each agent plays
the available strategy having highest virtual points; this allows an analytic
expression to be obtained for the probability that an agent plays a given
strategy for general $s$, which in turn yields 
\begin{equation} n_r=N\bigg(\bigg[1 - \frac{(r-1)}{2^{m+1}}\bigg]^s -
\bigg[1-\frac{r}{2^{m+1}}\bigg]^s\bigg)
\end{equation}
where $\sum_{r=1}^{2^{m+1}} n_r =N$ as required. Since agents are discrete objects, the simulation
tends to produce discrete steps in the curves of $n_r$ as a function of $r$. This
effect
becomes more pronounced as $m$ increases since the total number of
strategies $2^{m+1}$
then exceeds the population size $N$. We therefore convert the theoretical
$n_{r}$
values of Eq. (2) to an integer. For large $m$ such that $2^{m+1}\gg N$, the
resulting theoretical values are typically $n_r\sim 1$ for small $r$ and
$n_r=0$ for $r>N$. Figure 2 compares the
theoretical
values of $n_r$ calculated using Eq. (2) with $s=2$ and $N=101$, to
numerical values
taken from the MG simulation. The
agreement is good. 

For the
virtual-point ordered list $\{n_\rho\}$, the strategy
$\rho'=2^{m+1}+1-\rho$ is {\em always} anticorrelated to the strategy
$\rho$, i.e.
$\rho'\equiv \bar \rho$. Hence knowledge of the sort operation $\Gamma$
completely
determines where  each strategy's anticorrelated partner is located in the
popularity-ordered list $\{n_r\}$. Since we are here only interested
in time-averaged and run-averaged
$\sigma$, we only need to consider the probability distribution of locations of
$\bar r$ in the  popularity-ordered
list $\{n_r\}$. We
therefore replace the sort operation $\Gamma$ by a
probability function $P(r'={\bar r})$ which gives the
probability that any strategy $r'$ is the anti-correlated partner
of strategy $r$ in the list $\{n_r\}$.
Hence Eq. (1) becomes
\begin{equation}
\sigma =
\bigg[ \frac{1}{2} \sum_{r=1}^{2^{m+1}} \sum_{r'=1}^{2^{m+1}}
\frac{1}{4}|n_r-n_{r'}|^2 P(r'={\bar r})\bigg]^{\frac{1}{2}}
\
\
\end{equation}
where the factor
$\frac{1}{2}$ discounts double-counting.
There are two limiting cases. When the virtual-point ordered
list
$\{n_\rho\}$ and the popularity-ordered list $\{n_r\}$ are identical, then
$P(r'={\bar r})$ will be a $\delta$-function at $r'=2^{m+1}+1-r$ and hence
Eq. (3)
has the same form 
as Eq. (1). In the opposite case where the two ordered
lists
are uncorrelated, $P(r'={\bar r})$ should be a flat distribution. In each
case, Eqs.
(2) and (3) can be combined to obtain closed-form analytic solutions for
arbitrary $s$
and $N$.

Figure 3 shows $P(r'={\bar r})$ for $r=1$ as a function of $r'$, taken from
the
numerical MG  simulation at $m=2,5$ and $10$. For small $m$ ($m=2$) the
anticorrelated strategy to the most popular strategy (i.e. $r=1$) 
is at $r'=2^{m+1}$,
i.e. it is the least popular strategy. Hence $P(r'={\bar r})$ resembles the
$\delta$-function limiting case mentioned above. From Fig. 2 we know that
very few agents will therefore pick this anticorrelated strategy. Hence the
crowd-anticrowd
cancellation will be small and $\sigma$ will be large, as can be seen in
Fig. 1. As
$m$ increases ($m=5$) a remarkable effect occurs: the peak in $P(r'={\bar
r})$ moves
up toward $r=1$. Hence both $r=1$ and its anticorrelated partner $\bar r$
are now very
popular. Whereas for $m=2$ it seemed like there was an effective `repulsion'
between
$r$ and $\bar r$, for $m=5$ this now seems more like an attraction.
Amusingly, the
shape of $P(r'={\bar r})$ for $m=5$ is reminiscent of the screening effect
of a
negative charge cloud around a positive charge placed at $r=1$, or even a
bound
electron-hole pair (i.e. exciton) with the crowd (anticrowd) playing the
role of the
positive (negative) charge. For large $m$ ($m=10$), the ability of the
anticrowd to `screen' the crowd has decreased yielding a rather flat
distribution as
shown. The consequence of this strong crowd-anticrowd correlation which
appears as $m$
increases, is that the crowd and anticrowd become comparable in size. Hence
$\sigma$ is small for $m\sim 5-6$, in agreement with Fig. 1. 
Note that the MG cannot fully `optimize' itself by building equal-sized crowds and
anticrowds.  In modified models of MG\cite{prl}
where this in-built frustration in the strategy space is allowed to relax,
equal-sized
crowds and anti-crowds do naturally emerge.

We now consider analytic expressions for $\sigma$ for
general $s$ using Eqs. (2) and (3). For small $m$, 
the virtual-point ordered list $\{n_\rho\}$ and the popularity-ordered list
$\{n_r\}$
will be very similar, hence $P(r'={\bar r}) \sim \delta_{r',2^{m+1}+1-r}$. The
discreteness of the agents will be unimportant since $n_r\gg1$,
hence $n_r$ can
be treated as continuous. Equations (2) and (3) yield
\begin{eqnarray}
\sigma_{{\rm low}\ m} & = & \frac{N}{2}\bigg[ \sum_{r=1}^{2^{m}} \bigg[
\bigg(1-\frac{r-1}{2^{m+1}}\bigg)^s - \bigg(1-\frac{r}{2^{m+1}}\bigg)^s
\nonumber \\
& & - \bigg(\frac{r}{2^{m+1}}\bigg)^s + \bigg(\frac{r}{2^{m+1}}\bigg)^s
\bigg(1 -
\frac{1}{r}\bigg)^s
\bigg]^2\
\bigg]^{\frac{1}{2}}\ \ .
\end{eqnarray}
For $s=2$ this becomes
\begin{equation}
\sigma_{{\rm low}\ m} = \frac{N}{{\sqrt 3}\  2^{\frac{m}{2}+1}} \bigg[ 1 -
2^{-2(m+1)}\bigg] ^{\frac{1}{2}} \ \ .
\end{equation}
Figure 4 shows these analytic curves for $s=2$ and $s=4$ (solid
lines monotonically decreasing). As
might be expected using the extreme
$\delta$-function form for
$P(r'={\bar r})$, these curves are slightly higher than the numerical
results in Fig.
1 for small $m$. Now consider the opposite extreme of uncorrelated $r'$ and
$\bar r$,
i.e. the flat distribution
$P(r'={\bar r})\sim 2^{-(m+1)}$.
For $s=2$ this gives
\begin{equation}
\sigma_{{\rm low}\ m} = \frac{N}{{\sqrt 3}\  2^{\frac{(m+3)}{2}}} \bigg[ 1 -
2^{-2(m+1)}\bigg] ^{\frac{1}{2}} \ \ .
\end{equation}
Equation (6) typically produces lower estimates for each $s$ value
(dashed lines). Values of $\sigma$ obtained from separate numerical
runs tend to be scattered in the region of these curves. For
larger $m$
($m>6$) we cannot ignore the discreteness of the agents 
(Fig. 2).
In this regime, $n_r\sim 1$ for $r<N$ while $n_r=0$ for $r>N$. Using the integer
form of Eq. (2) for high $m$, and the flat distribution
for $P(r'={\bar r})$, yields
\begin{equation}
\sigma_{{\rm high}\ m} =  \frac{\sqrt N}{2} \bigg[ 1 -
\frac{N}{2^{m+1}}\bigg]^{\frac{1}{2}}\ .
\end{equation}
This expression approaches the coin-toss limit from below as $m\rightarrow
\infty$, as shown in Fig. 4 (solid line monotonically increasing).
Within the approximation used
here, this curve is insensitive to $s$. Note that the numerical MG
results are also
consistent with this finding of weak $s$-dependence for large $m$.
Hence Fig. 4 gives
a clear picture of what happens to $\sigma$ as $s$ 
and $m$ increase: considering
the
monotonically decreasing curve for low $m$, and the monotonically increasing
curve for
high $m$, we see that (a) there should be a minimum in $\sigma$ for $s=2$,
(b) this
minimum should move to higher
$m$ as
$s$ increases and (c) the minimum should become shallower as $s$ increases.
Each of
these statements agrees with numerical simulation results (c.f.
Fig. 1).
In addition, the curve $\sigma_{{\rm high}\ m}\rightarrow 0$ at $N=2^{m+1}$,
i.e. for $m\sim 5-6$.

We now compare the theoretical crowd-anticrowd calculation (Eq. (3)) with the
numerical simulation. The most interesting case is $s=2$. Figure 5 shows the
spread of numerical values for different runs (open circles) compared to
theory (solid circles). The
agreement is good. The appropriate analytic expressions for the
probability
function
$P(r'={\bar r})$ in Eq. (3) involve multiple sums and are complicated:
we therefore obtained the results for each $m$ in Fig. 5 by generating the
corresponding 
$P(r'={\bar r})$ forms to those in Fig. 3. The theoretical
points tend to lie in between the limiting curves of Fig. 4 {\em except} at the
minimum where the
remarkable form for $P(r'={\bar r})$ (recall Fig. 3) pulls the theoretical
value slightly
below the cruder limiting curves of Fig. 4. Elsewhere we will discuss
simpler closed-form expressions for
$P(r'={\bar r})$, together with 
an analysis of the time-series $A(t)$ and history bit-string
statistics within the crowd-anticrowd model \cite{long}.

In summary we presented an analytic analysis of crowding
effects in MG which offers a novel explanation of the
main finding of Ref.
\cite{savit}. Future work 
will explore possible connections to the spin-glass formalism
of  Challet {\em et al} \cite{spinglass}.

We thank T.S. Lo, J.P. Garrahan, D. Sherrington and D. Challet for
discussions.

\begin{figure}
\bigskip

\caption{Standard deviation $\sigma$ for the Minority Game as a function of
memory-size $m$
for $s=2,3,4$ strategies per agent and $N=101$ agents. Solid curve:
numerical
simulation. Dashed curve: crowd-anticrowd theory using Eq.
(1). Random (coin-toss) limit $\sigma={\sqrt N}/2=5.0$ is indicated.}
\bigskip
\bigskip

\caption{Histograms of the number of agents using strategy $r$
for $r=1$ (most popular) to $r=2^{m+1}$ (least
popular). Results are shown for $m=2$ (left scale) and $m=7$ (right scale).
Solid lines: numerical simulation. Dashed lines: theory from Eq. (2).}
\bigskip
\bigskip

\caption{Probability function $P(r'={\bar r})$ giving the probability that the
strategy
ranked $r'$ on the popularity-ordered list, is anti-correlated with the
strategy
ranked $r$. Results are shown for $r=1$ (i.e. most popular strategy) as a
function of
$r'$ for
$m=2$ (dotted-dashed), $m=5$ (dotted) and
$m=10$ (solid).
$s=2$ and $N=101$. Note that $\sum_{r'} P(r'={\bar r}) =1$.}
\bigskip
\bigskip

\caption{Theoretical curves for $\sigma$ using Eqs.
(4)-(7). Monotonically decreasing curves for $s=2,4$ at low $m$ ($m<6$): solid
lines correspond to
$\delta$-function
$P(r'={\bar r})$ distribution 
neglecting agent discreteness, dashed lines correspond to flat distribution
neglecting agent discreteness. Monotonically increasing solid line for large $m$
($m>6$) is  independent of $s$: it corresponds
to a flat distribution and accounts for agent discreteness.}
\bigskip
\bigskip

\caption{Theoretical crowd-anticrowd calculation (solid circles) vs. numerical 
simulations (open circles)
for
$s=2$, $N=101$.
16 numerical runs are shown for each $m$.}

\end{figure}

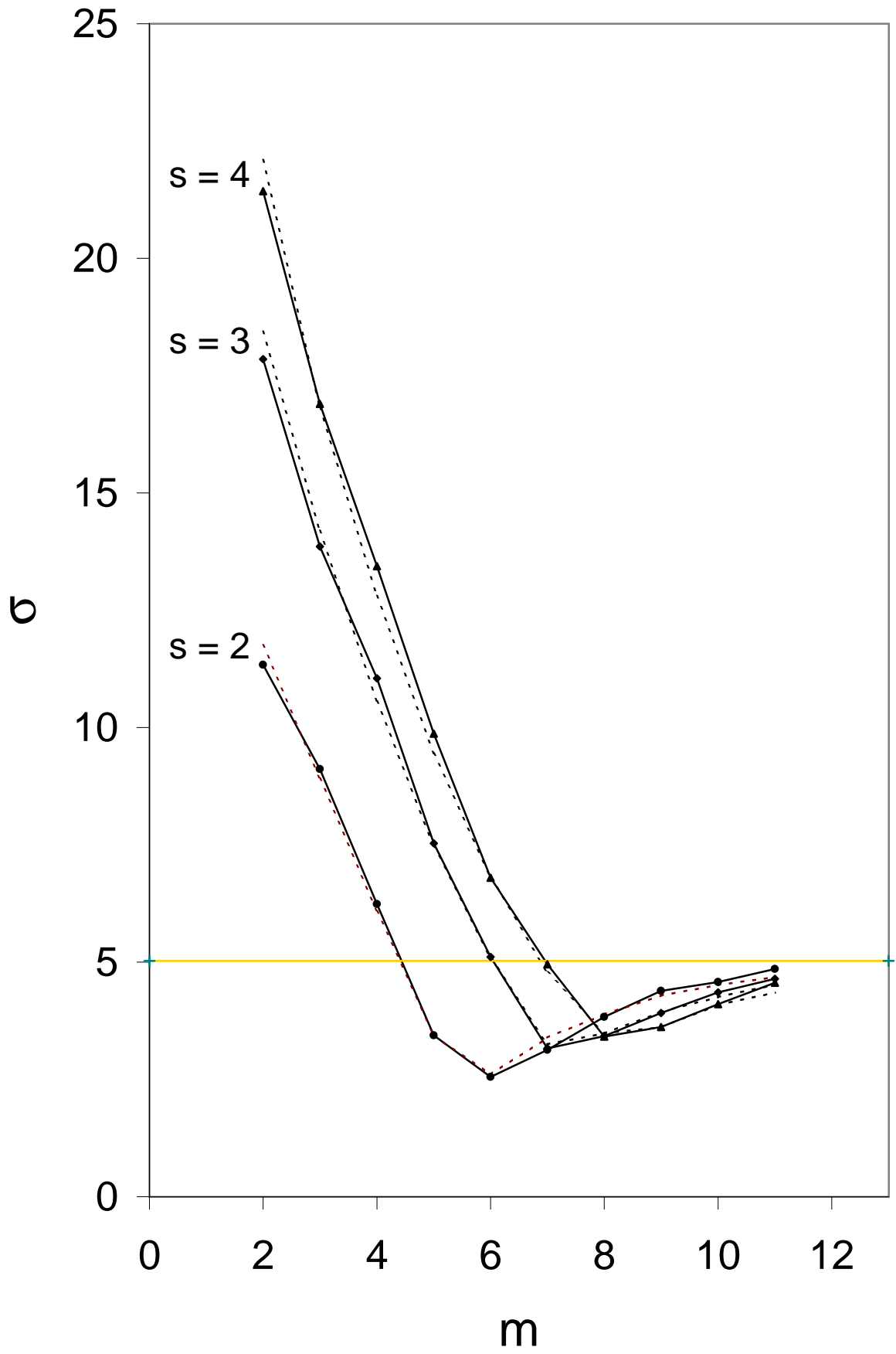
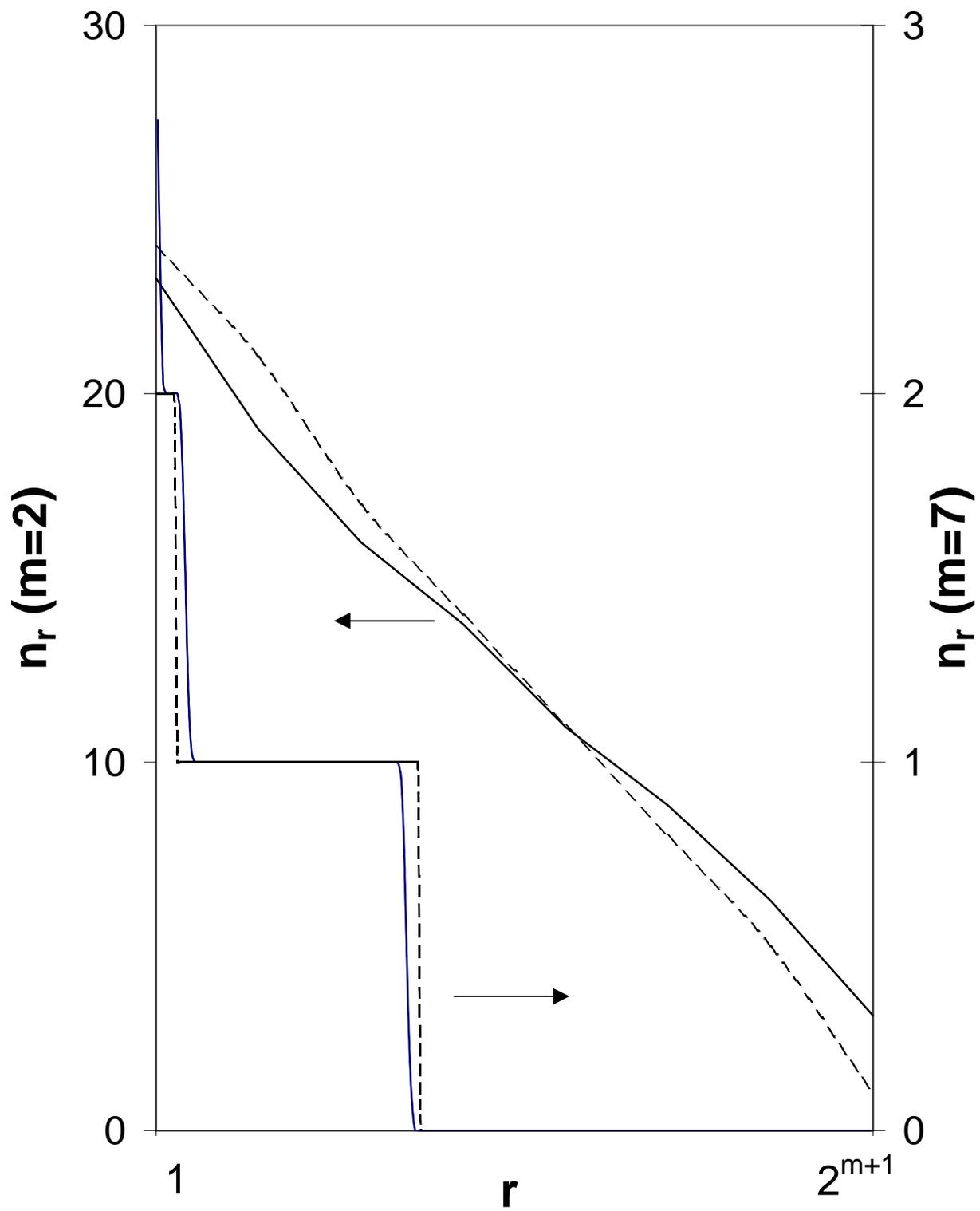
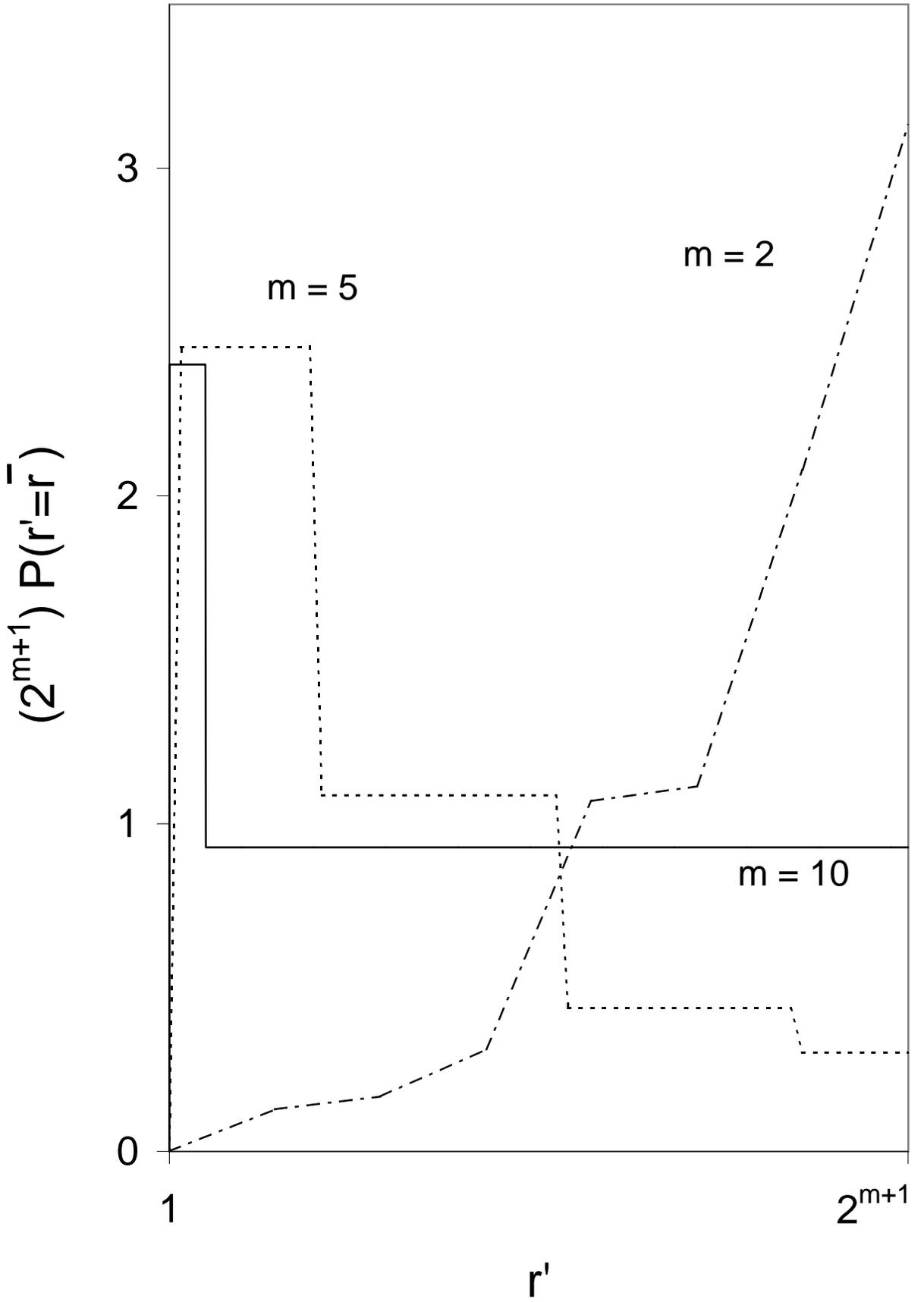
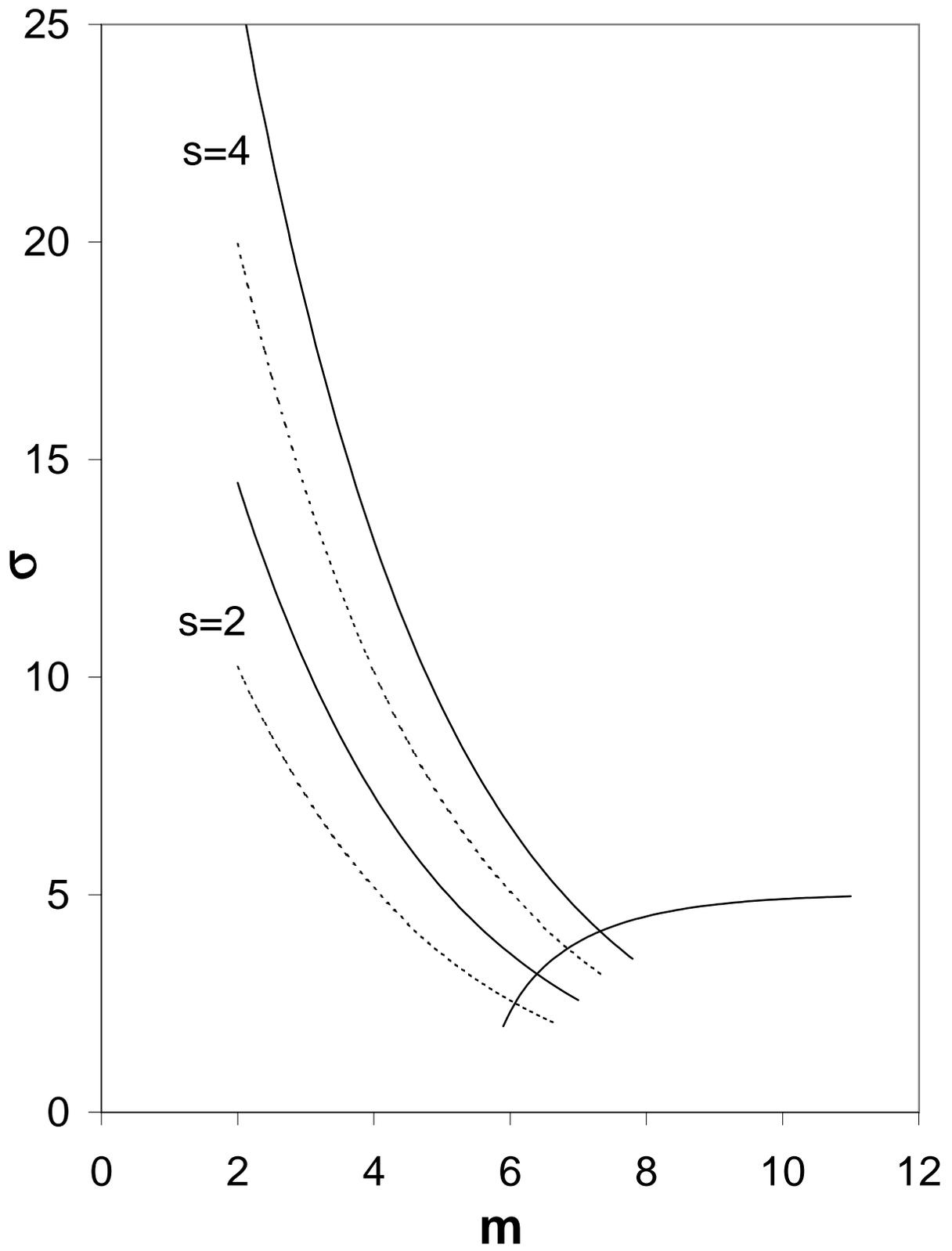
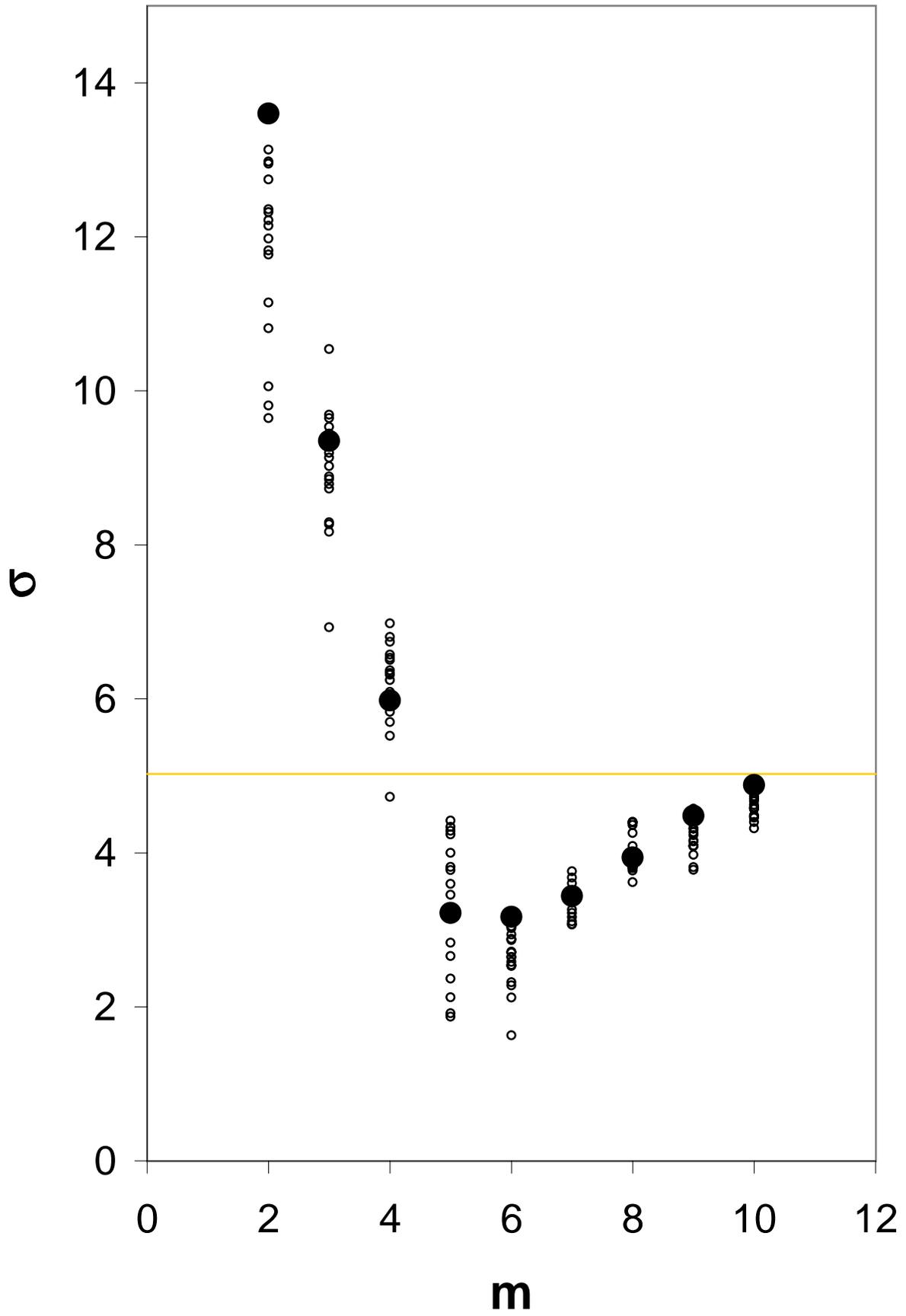
\end{document}